# Practical Limits of Achieving Artificial Magnetism and Effective Optical Medium by Using Self-Assembly of Metallic Colloidal Clusters

*Kwangjin Kim*[1,2,+], *Ji-Hyeok Huh*[1,2,+], *Doyoung Yu*[1,2,+], and *Seungwoo Lee*[2*]

[1]SKKU Advanced Institute of Nanotechnology (SAINT), Sungkyunkwan University, Suwon 16419, Republic of Korea
[2]KU-KIST Graduate School of Converging Science and Technology, Korea University, Seoul 02841, Republic of Korea

[*]Email: seungwoo@korea.ac.kr
[+]Equally contributed to this work

Keywords: Colloids, Self-Assembly, Plasmonic Metamolecules, Magnetism

Abstract: The self-assembly of metallic colloidal clusters (so called plasmonic metamolecules) has been viewed as a versatile, but highly effective approach for the materialization of the metamaterials exhibiting artificial magnetism at optical frequencies (including visible and near infrared (NIR) regimes). Indeed, several proofs of concepts of plasmonic metamolecules have been successfully demonstrated in both theoretical and experimental ways. Nevertheless, this self-assembly strategy has barely been used and still remains an underutilized method. For example, the self-assembly and optical utilization of the plasmonic metamolecules have been limited to the discrete unit of the structure; the materialization of effective optical medium made of plasmonic metamolecules is highly challenging. In this work, we theoretically exploited the practical limits of self-assembly technology for the fabrication of optical magnetic metamaterials.

TOC:

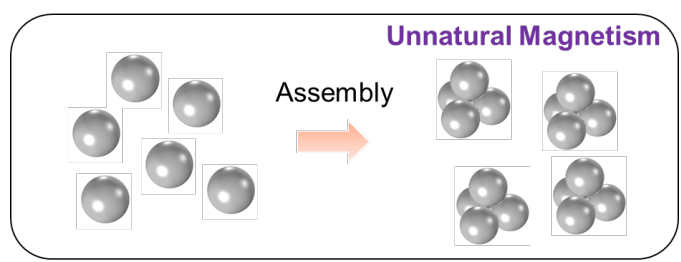

## 1. Introduction

By taking advantages of recent advances in chemical synthesis, variously shaped metallic colloids ranging from 5~200 nm in size have become accessible with high yield;[1-4] we can use these subwavelength-scale colloids as optical meta-atoms with a wide range of electric and magnetic resonance tunability.[5-18] For example, the closely-packed metallic colloidal clusters can form a ring inclusion with a nanometer gap in order to boost the circulating displacement current and the resultant optical magnetism.[6,9-12] This artificial magnetism is essential for the unnatural reduction of the effective permeability ($\mu$) and $n$.[5-18]

While they provide opportunities for achieving the unnaturally low refractive indices at optical frequencies, colloidal assemblies are still considered an underutilized approach in optical magnetic metamaterial research. In this article, we address why colloidal self-assembly has not yet been fully generalized for the materialization of optical magnetic metamaterials. By numerically exploiting colloidal assemblies as the building blocks of optical magnetic metamaterials, we can systematically verify their practical limitations in terms of achieving unnatural refractive index.

## 2. Methods

### 2.1. Numerical simulations

The finite element method (FEM), supported by home-built code, was used for the numerical calculation of electric and magnetic resonances including electric/magnetic dipolar (ED/MD) modes. The complex dielectric constant of silver (Ag) was taken from the Johnson and Cristy library.[19] In order to exploit the fundamental limit of unnatural refractive index for optical metafluids, we used silver (Ag) as an element of plasmonic metamolecules rather than some other noble metals (e.g., gold (Au)). The multipole expansions were used to numerically calculate the ED/MD scattering cross section (SCS, nm$^2$).[20] First, the induced electric (**p**) and magnetic (**m**) dipolar moments of the colloids were obtained through the following equations:

$$\mathbf{p} = \int \varepsilon_0 (\varepsilon - 1) \mathbf{E} dr \tag{1}$$

$$\mathbf{m} = \frac{-i\omega}{2} \int \varepsilon_0 (\varepsilon - 1) [\mathbf{r} \times \mathbf{E}] dr \tag{2}$$

where $\varepsilon_0$ and $\varepsilon$ are respectively permittivities in air and host medium. $E$ indicates the electric field in metals. Thus, ED SCS ($W_p$) and MD SCS ($W_m$) are expressed as:

$$W_p = \frac{k_0^4}{6\pi \varepsilon_0^2 E_0^2} |\mathbf{p}|^2 \tag{3}$$

$$W_m = \frac{\eta_0^2 k_0^4}{6\pi E_0^2} |\mathbf{m}|^2 \tag{4}$$

where $k_0$, $\eta_0$, and $E_0$ indicate wavenumber and impedance of light in host medium and electric field in vacuum.

We used finite-difference, time-domain (FDTD) to numerically calculate the scattering-parameters, which were used to obtain the effective parameters of the regularly arrayed, 2D colloidal crystals (i.e., high-index optical metamaterials). The gap was set at 1 nm between each metallic colloid. In addition, Ag was used as an element of metallic colloids.

### 2.2. Effective medium theory with 3D dressed polarizability

The Maxwell-Garnett (M-G) relation provides the rationalization for the effective parameters of the medium, where colloids are randomly distributed (less than 50 vol%).[6,9,14] Put simply, the electric and magnetic polarizabilities from each meta-atoms (or colloids) can be abstracted by the FEM method; then, the effective parameters (i.e., effective permittivity ($\varepsilon_{eff}$) and effective permeability ($\mu_{eff}$)) can be calculated by the following M-G relations:

$$\frac{\varepsilon_{eff}-1}{\varepsilon_{eff}+2}=\frac{N\alpha_e}{3} \qquad (5)$$

$$\frac{\mu_{eff}-1}{\mu_{eff}+2}=\frac{N\mu_0\alpha_m}{3} \qquad (6)$$

where $N$, $\alpha_e$, and $\alpha_m$ are the number density, electric polarizability and magnetic polarizability of metamolecules.

Furthermore, when the density of colloids is increased (more than 50 vol%), then the electric and magnetic interactions between colloids must be considered. Thus, we employed dressed polarizability theory to the calculate effective parameters in this case.[6,21] From dressed polarizability theory, the localized electric ($\mathbf{E}_{loc}$) and magnetic field ($\mathbf{H}_{loc}$) which are affected by the nearby scattering of colloids can be calculated using Green's function (**G**) and the following equations:

$$\mathbf{E}_{loc}=\mathbf{E}_0+\omega^2\mu_0\sum\mathbf{Gp} \qquad (7)$$

$$\mathbf{H}_{loc}=\mathbf{H}_0+\omega^2\varepsilon_0\mu_0\sum\mathbf{Gm} \qquad (8)$$

where $\mu_o$ is permeability in free space. Thus, we obtained dressed polarizability ($\alpha_{e,dressed}$, $\alpha_{m,dressed}$) as the following relation:

$$\mathbf{p}=\varepsilon_0\alpha_{e,dressed}\mathbf{E}_{loc} \qquad (9)$$

$$\mathbf{m}=\mu_0\alpha_{m,dressed}\mathbf{H}_{loc} \qquad (10)$$

Finally, we can estimate the effective parameters with consideration of the interaction between colloids from the M-G relation through the use of dressed polarizability rather than original polarizability.

3. Result and Discussion

3.1. Design validity of metallic colloidal clusters for artificial magnetism.

Assembling deep-subwavelength scaled, several Ag nanospheres (NSs, ~ less than 60 nm) into colloidal clusters can form a ring inclusion geometry, which can induce a sufficiently strong circulating displacement current, resulting in an induced magnetic dipole (i.e., magnetism) in the visible regime.[7-10,14-16] For example, as shown in **Figure 1a**, a 3D symmetric tetrahedral cluster consisting of four 60 nm-sized Ag NSs with a 1 nm gap can induce artificial magnetism (magnetic dipolar (MD) resonance), as evidenced by a distinct scattering spectral shoulder (i.e., scattering cross section (SCS, nm$^2$)) at the 685 nm wavelength. In contrast, the counterpart without any gaps (i.e., where Ag NSs are directly connected with each other by an Ag nanorim) exhibited negligible MD resonance (see **Figure 1b**); only the scattering peak corresponding to the electric dipolar (ED) resonance was visible. This result implies that the capacitive coupling between Ag NSs plays a pivotal role in boosting the circulating displacement current. The strength of a capacitive coupling is generally enhanced with reducing the gaps between metallic units.

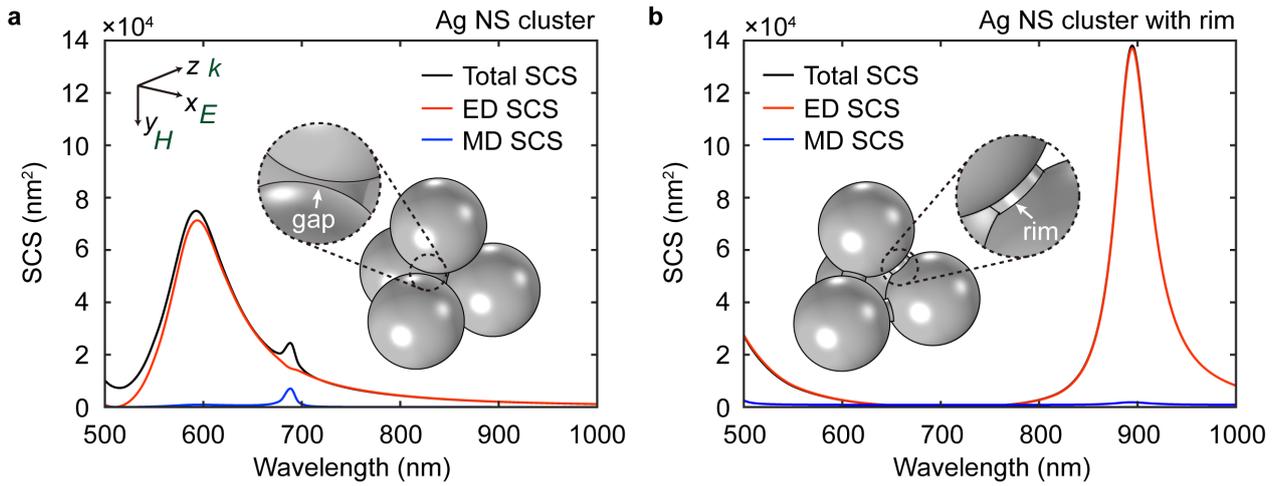

**Figure 1.** Scattering cross section (SCS) of silver nanosphere (Ag NS) tetrahedral cluster without (a) and with (b) Ag rim. 60 nm Ag NSs are assumed to be clustered into a tetrahedron with 1 nm gaps. SCS was numerically calculated by the finite-element method.

3.2. Fundamental limitations of index-engineering using plasmonic metamolecules.

Such Ag tetrahedral clusters serving as plasmonic metamolecules are still in the deep-subwavelength scale (less than 120 nm) compared to the visible wavelengths; thus, Ag tetrahedral clusters-dispersed fluids themselves can be considered optical metamaterials. This soft fluidic platform of metamaterials can be categorized into a distinct concept of "optical metafluids".[6,11,14] When the volume fraction is less than 50 %, plasmonic metamolecules can be randomly and homogeneously dispersed within the host fluids;[6,22] thus, the M-G relation (or Clausius-Mossotti relation) can be used to obtain the effective parameters (i.e., effective $\varepsilon$, $\mu$, and $n$).[6] In the M-G relation, the effective parameters can be abstracted out by the total summations of the electric and magnetic polarizabilities, which are accessible from each individual meta-atom or metamolecule. Herein, these polarizabilities were obtained through numerical simulation (FEM, as discussed in the Methods section).

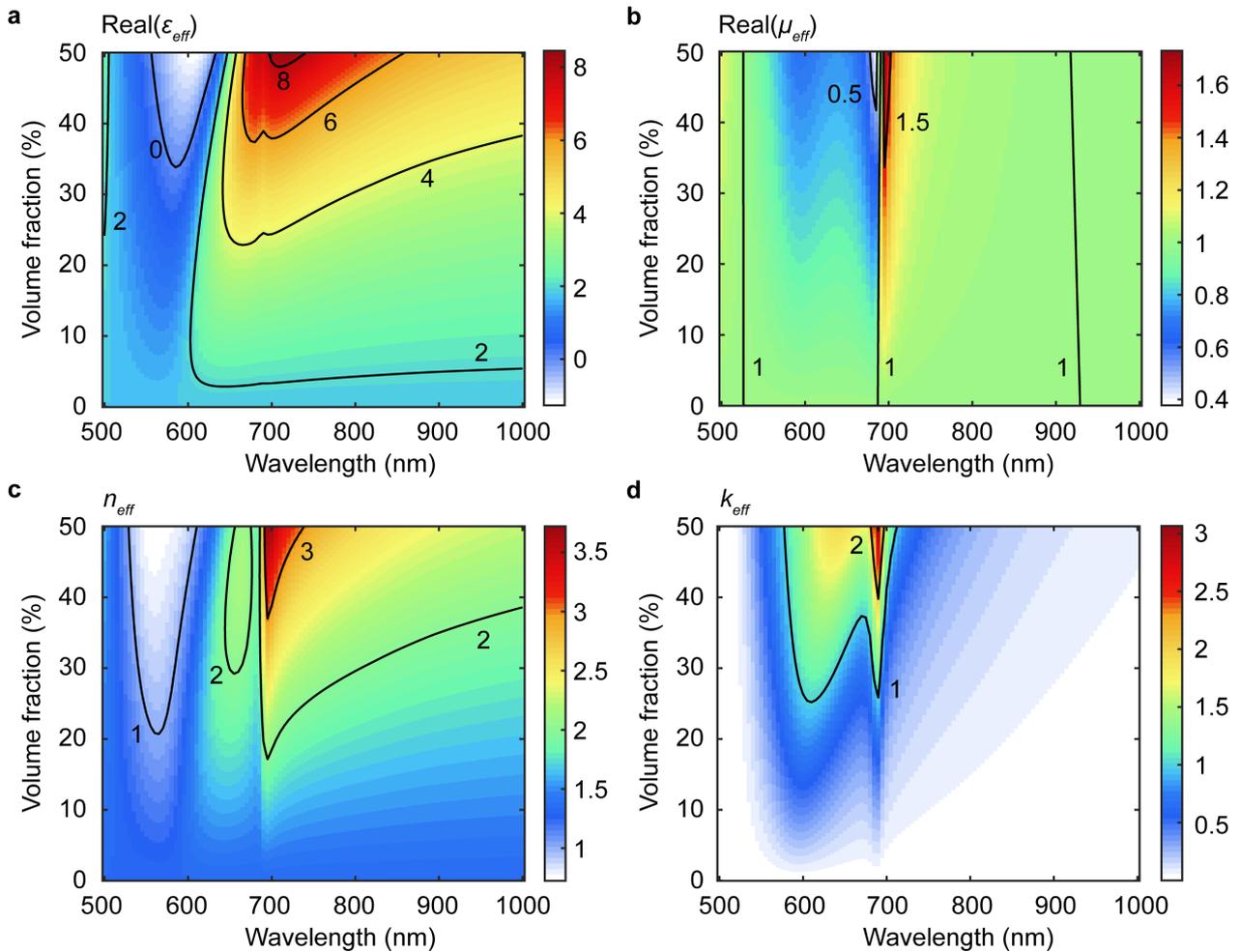

**Figure 2.** Effective parameters, which are fundamentally accessible with optical metafluids: (a) real part of permittivity (ε), (b) real part of permeability (μ), and (c) real and (d) imaginary parts of refractive index (n). Herein, optical metafluids consist of (i) a tetrahedral cluster (symmetrically clustered 60 nm Ag NSs with 1 nm gaps) and (ii) water host medium. Such a collective set of effective parameters was exploited as a function of metamolecular volume % (vol %).

**Figure 2** summarizes the available range of effective parameters according to the volume fraction (vol %) of Ag tetrahedral clusters (plasmonic metamolecules) within water (host medium). Herein, Ag NSs within clusters were assumed to be separated by 1 nm gaps, as shown in **Figure 1a**. For gaps smaller than 1 nm, quantum behavior (i.e., non-local effects) can occur,[23] and thus the capacitive coupling between Ag NSs can be dimmed. Both ε and μ of optical metafluids were modulated according to the Lorentzian oscillator model.[6] As expected, ε can be negative at a relatively high volume fraction of Ag tetrahedral clusters (**Figure 2a**). However, ~ 0.6 of μ was the lower limit, accessible with a random dispersion of Ag tetrahedral clusters (**Figure 2b**). Thereby, negative $n$ cannot be achieved, as shown in **Figure 2c** (0.8 of $n$ was a lower limit). Substantial optical loss is another drawback, which is inevitable from the use of plasmonic metamolecules (**Figure 2d**).

A. Alù and colleague originally argued that 74 vol% of 2D tetrameric plasmonic metamolecules (76 nm Ag NSs) within in an air host medium can lead to an unnatural negative $n$.[9] This result was obtained using the M-G relation. However, using the M-G relation to calculate effective parameters of such an optical effective medium is not justifiable for a realistic condition. The value of 74 is known as the maximal vol % of nanoparticles for spherical colloidal crystals (i.e., face-centered cubic (FCC) lattice). The plasmonic metamolecules (especially 3D clusters) can also be approximated by spherical colloidal particles. For instance, a polymeric encapsulation can smooth the protrude morphology of plasmonic metamolecules into a spherical morphology.[6] Therefore, the optical effective medium can in principle be filled with 74 vol % of plasmonic metamolecules. However, it is important to note that the induced ED and MD per each plasmonic metamolecule, crystalized into an FCC lattice, should strongly interact with each other (e.g., in the form of capacitive coupling between the EDs of each metamolecule). Unfortunately, the M-G relation cannot account for this interaction.

In order to address this challenge, we advanced the effective medium theory for colloidal "superlattice" metamaterials by which the 3D dressing of electric and magnetic polarizabilities rationalizes the interactions between the induced dipoles (i.e., ED and MD).[6] **Figure 3** shows the calculated effective parameters for the FCC crystal of 74 vol% Ag tetrahedral clusters (composed as in **Figure 2**). In order to elucidate the importance of the induced dipolar interactions, the results calculated using the M-G relation, which is modified with 3D dressed polarizabilities (right panels (**Figures 3b, d, f**)), were compared with those obtained from the bare M-G relation (left panels (**Figures 3a, c, e**)). In order to reflect a realistic case of optical metafluids, we used water as a host medium rather than air. The bare M-G relation resulted in a $\mu$ of almost zero at around the wavelength of magnetic resonance; as a result, $n$ can be reduced to near zero at the wavelength of 690 nm. If we use air as a host medium, $\mu$ and $n$ could become unnaturally negative.

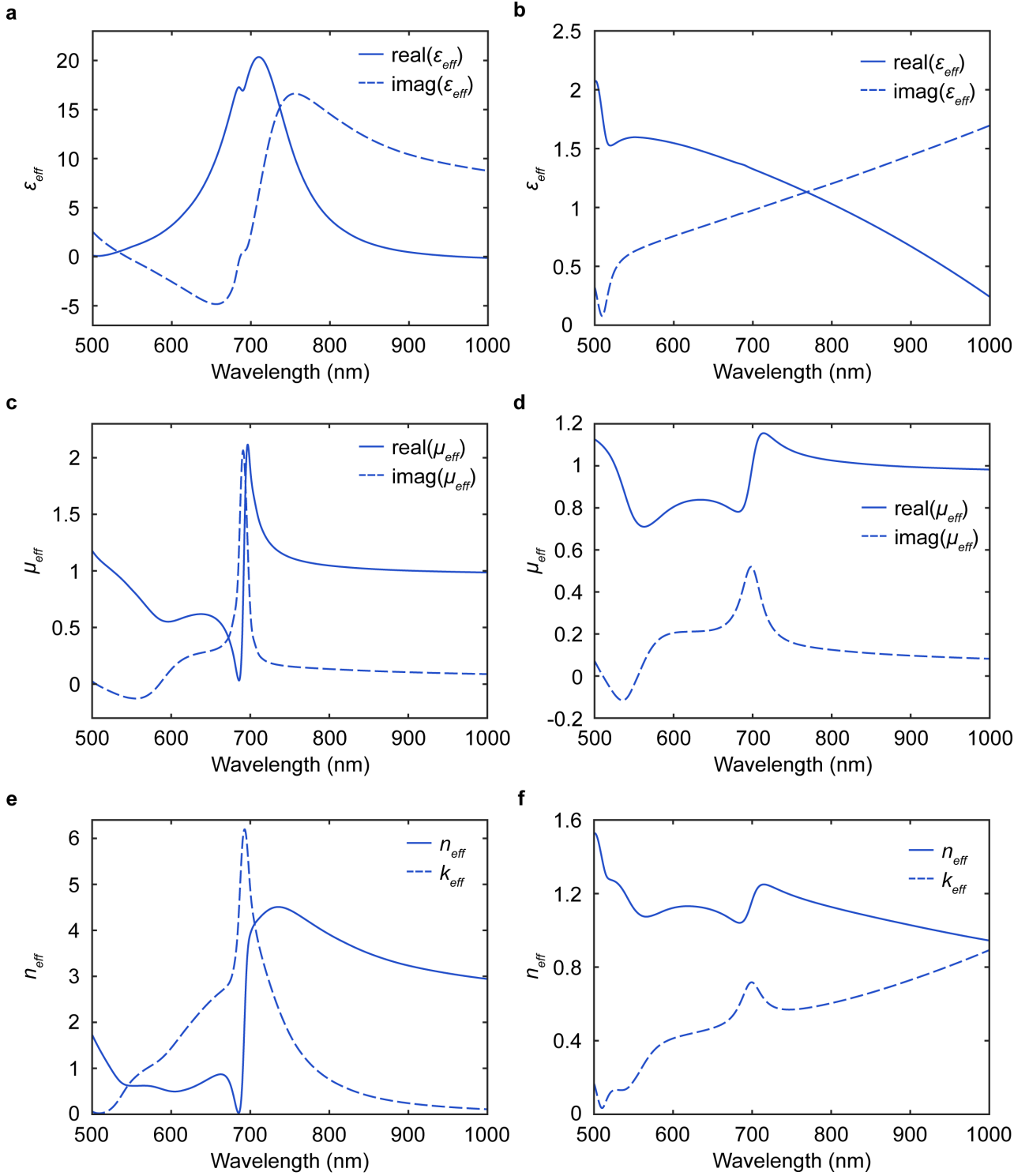

**Figure 3.** Effective parameters of plasmonic metamolecular crystals (i.e., face-centered-cubic (FCC) crystals) with a 74 volume fraction (vol %), which were calculated by the Maxwell-Garnett (M-G) relation without (a, c, e) and with (b, d, f) modulation using 3D dressed polarizabilities. Herein, plasmonic metamolecules were composed as in Figure 1a and Figure 2.

However, the 3D dressed polarizabilities in the M-G relation resulted in *n* values of 0.9 ~ 1.5 at the wavelength of interest, which is a naturally accessible regime. For the case of ED, a strong capacitive coupling can occur between each plasmonic metamolecule; consequently, the resonance wavelength of ED was significantly red-shifted. In contrast, the MD interactions were not influenced by capacitive coupling, as evidenced by the fact that their resonance wavelength remained intact as compared to that of the individual plasmonic metamolecules. Also, the strength of MD resonance was weakened following the implementation of the 3D dressed polarization into the M-G relation. Overall, at the visible regime, *n* was negligibly modulated in the FCC crystal of 74 vol% Ag tetrahedral clusters.

3.3. Practical limitations on the reliable self-assembly of plasmonic metamolecules

Another limitation of magnetic optical metafluids is that ED and MD resonances could be heavily influenced even by just a few nanometer-structural errors, which are vulnerable to the self-assembly of colloidal clusters (e.g., convective forcing and DNA complementary binding). For example, two gaps of 3D tetrahedral clusters composed of 60 nm Ag NSs could be deviated from 1 nm to 2 nm in a realistic experiment of self-assembly; this leads to the slight breaking of the symmetry of 3D plasmonic metamolecules. Interestingly, it turned out that this slight breaking of the symmetry of the cluster boosted the strength of the induced MD, as shown in **Figure 4a**. This occurred because the energy of the induced ED was partially transferred to the induced MD, as evidenced by the dip of ED SCS (**Figure 4b**). Note that the wavelength of this ED SCS dip matched that of the MD SCS peak (635 nm wavelength). This energy transfer between ED and MD is referred to as Fano resonance (magnetic Fano resonance).[12,24,25]

In a symmetric cluster, ED and MD cannot interact with each other; thus, the SCS peak (605 nm wavelength) and shoulder (690 nm wavelength), respectively corresponding to the ED and MD resonances, were independently observed without any interferences. This calculation result implies that the resonant behaviors of plasmonic metamolecules could be dramatically changed even with a 1 nm gap error. Therefore, it is difficult to assure the reliable materializations of plasmonic metamolecules through self-assembly. Recently, atomic force microscopy (AFM)-enabled manipulation of metallic colloids has been used to assemble plasmonic metamolecules.[12,25] The kicking and dribbling of individual metallic colloids by the manipulation of AFM tip with a nanometer resolution enable the on-demand assembly of 2D and 3D plasmonic metamolecules;[12] however, it is still prone to the nanometer scale error.

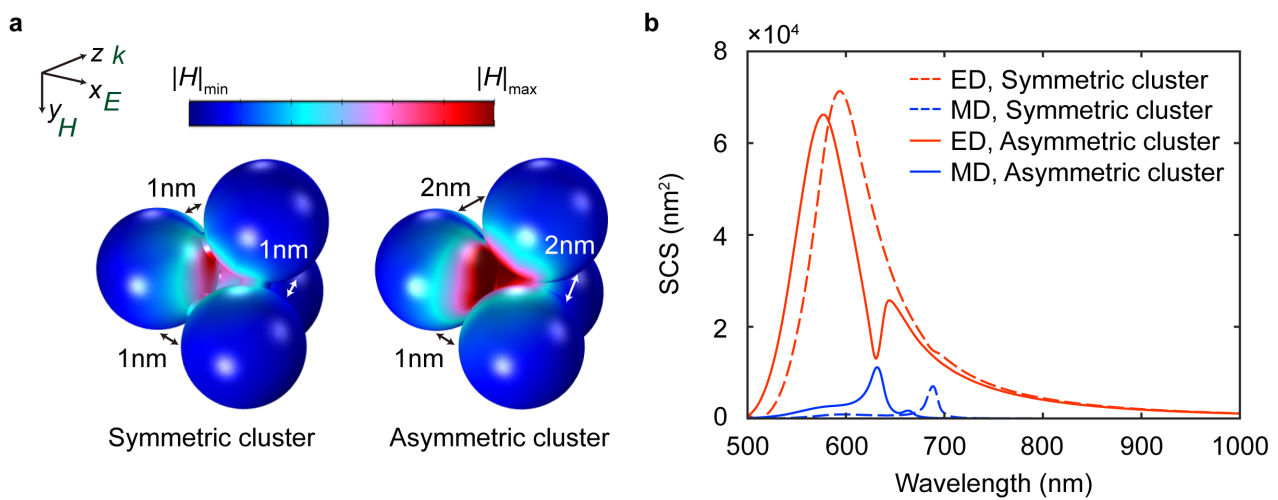

**Figure 4.** The influence of the 1 nm structural error (gap error) on the magnetic resonances: (a) spatial mapping of the induced magnetic field intensity and (b) SCS.

More critically, the resonant behaviors of asymmetric 3D plasmonic metamolecules should depend on the angle and polarization of the incident light. This means that the ED and MD resonances of each 3D asymmetric plasmonic metamolecule, self-assembled from metallic colloids, are likely changed during Brownian motions in the fluidic medium. Thus, in a realistic situation, the effective parameters of optical metafluids, which result from the ensemble of whole 3D plasmonic metamolecules dispersed in a host fluid, would not be comparable to the fundamental limits defined above.

4. Conclusions.

In summary, we theoretically exploited the design rule of metallic colloidal magnetic metamaterials by numerical simulation in conjunction with effective medium theory. The self-assembly of metallic colloidal clusters can in principle provide a versatile, but highly effective platform toward both unnaturally low refractive indexes at optical frequencies. However, several experimental limitations were also theoretically determined, which are vulnerable to a realistic situation and could make the self-assembly of metallic colloids an underutilized toolset, particularly for artificial magnetism.

Acknowledgement

This work was supported by Samsung Research Funding Center for Samsung Electronics under Project Number SRFC-MA1402-09.